\documentstyle[11pt,newpasp,twoside]{article}
\def\edcomment#1{\iffalse\marginpar{\raggedright\sl#1\/}\else\relax\fi}
\reversemarginpar

\begin{document}
\title{Secular Evolution of Disks}
\author{Xiaolei Zhang}
\affil{Raytheon ITSS/NASA GSFC,
Code 685, Greenbelt, MD 20771}

\begin{abstract}
It is found that a previously
thought-to-be well established result of density wave theory, that there is
no interaction between a quasi-stationary spiral density wave and the
basic state (i.e. the axisymmetric part) of the galactic disk, 
is in fact false. When solved as a nonlinear and globally
self-consistent problem, the presence of non-axisymmetric patterns
such as spirals or bars is shown to lead to significant interaction 
of the basic state and the density wave, with the net result being the
simultaneous acquirement of a quasi-steady wave amplitude and the secular
redistribution of disk matter.  The secular operation of this
dynamical mechanism leads to the evolution of the Hubble type
of a galaxy from late to early.
\end{abstract}

\section{Introduction}

During the past decade, growing evidence has pointed to a trend
of secular morphological evolution of galaxies over the cosmic time.
A firm theoretical foundation needs to be established on the
operations of the dominant mechanisms driving this secular evolution.
In what follows, we present a dynamical mechanism responsible
for the secular morphological evolution of galaxies which is {\em
internal} to galaxies which possess large scale spiral or bar patterns.
We demonstrate the validity and inevitability of the
operation of this mechanism, the astrophysical consequences
as well as the 
connection of the current work with previous work in this field.

\section{Basic Results}

In was first shown in Zhang (1996) that for a spontaneous spiral mode,
the potential spiral lags the density spiral in phase inside corotation,
and vice versa outside corotation.  The phase shift between the potential 
and density spirals means that there is a torque exerted by the 
potential spiral on the density spiral, and a secular transfer 
of energy and angular momentum between the disk matter and
the density wave at the quasi-steady state of the wave mode (Zhang 1998,1999).  
The torque T(r) applied by the spiral potential on the density
in an annulus of unit width can be written as
$$
T (r) = dL/dt
= r 
\int_0^{2  \pi}
- \Sigma (r \times \nabla {\cal{V}})_z d \phi
$$
\begin{equation}
=
r \int_0^{2 \pi} - \Sigma_1 { {\partial {{\cal{V}}_1}} \over {\partial \phi}}
d \phi
=- \pi m r \Sigma_1(r) {\cal{V}}_1(r) \cdot \sin (m \phi_0(r))
,
\end{equation}
where $\Sigma$, ${\cal{V}}$, $\Sigma_1$, ${\cal{V}}_1$ 
are the surface density and potential,
as well as the spiral perturbation density and potential in the annulus,
respectively, $L$ is the angular momentum of the disk matter in the annulus, 
$\phi_0$ is the potential-density phase shift and $m$ is the number of
spiral arms. 

At the quasi-steady state, the energy and angular momentum 
transfer between the basic state matter and the spiral density wave
is achieved through a temporary local 
gravitational instability at the spiral arms (Zhang 1996). The 
length scale of this instability at the solar neighborhood is about 1 kpc,
which coincides with the length scale of the giant
molecular and HI complexes near the Galactic spiral arm region.
The presence of the instability condition at the spiral arms,
coupled with the supersonic to subsonic transition of particle
streaming velocity with respect to the spiral arm, indicate that
the nature of the large-scale spiral pattern in galaxies is in
fact spiral gravitational shocks.

\section{Astrophysical Consequences}

The wave-basic state interaction leads naturally to the damping
of the growing wave mode and to to the acquirement of quasi-steady state
(Zhang 1998).  A by-product of the wave-basic state interaction is that 
an averge star inside corotation will tend to
lose energy and angular momentum to the wave secularly
and spiral inward.  The rate of this orbital delay can be shown 
to be 
\begin{equation}
{{dr} \over {dt}}
= - { 1 \over 2} F^2 v_0 \tan i \sin(m \phi_0)
\end{equation}
where $F$ is the fractional wave amplitude, and $v_0$ is the circular
velocity of the star, and $i$ is the pitch angle of the spiral.
This evolution rate expression
has been quantitatively confirmed in the N-body
simulations (Zhang 1998).
This orbital decay rate is
about 2 kpc per $10^{10}$ years for our own Galaxy, which
corresponds to a mass accretion rate of about $6 \times 10^9 M_{\odot}$
per $10^{10}$ years.  A substantial fraction of the bulge can thus
be built up in a Hubble time.

Another important consequence of spiral-induced wave-basic state
interaction is the secular heating of the disk stars.
Since a spiral density wave of pattern speed $\Omega_p$
can only gain energy and angular momentum
in proportion to $\Omega_p$, the pattern speed of the wave,
and a disk star which moves on a nearly circular
orbit loses its orbital energy and angular momentum
in proportion to $\Omega$, the circular speed of the star,
an average star cannot lose
orbital energy entirely to the wave for galactic radii other
than the corotation, and thus the excess energy
is used for the secular heating of the disk stars.  For our Galaxy,
the diffusion coefficient due to the spiral-induced secular heating is
estimated to be
\begin{equation}
D = (\Omega - \Omega_p) F^2 v_c^2 \tan i \sin(m \phi_0 ) \approx
6.0 (km s^{-1})^2 yr^{-1},
\end{equation}
if using the same set of spiral parameters
as used above for estimating Bulge building.
This value of D fits very well the age-velocity
dispersion relation for the solar neighborhood stars (Zhang 1999).
The above expression for D can be shown to be approximately constant
across the galactic radii (Zhang 1999), which agrees with the known
isothermal distribution of the stellar and gaseous mass.
Similar energy injection into the interstellar medium can serve
as the top-level source for the subsequent supersonic turbulence cascade.

In general, the radial mass accretion process causes the disk mass
to be more and more centrally concentrated, and causes the morphological
type of a galaxy to evolve from late to early along the Hubble sequence.
Such morphological transformation is most pronounced in dense clusters,
which is the well-known Butcher-Oemler effect.  In the current
scenario, the enhanced mass accretion for cluster galaxies is due
to the large amplitude and open spiral patterns induced through
tidal interactions among cluster members, since the effective evolution
rate due to spiral structure is seen to be proportional to
wave amplitude squared and the spiral pitch angle squared
(equation 2, note that the
phase shift $\phi_0$ itself is approximately proportional 
to spiral pitch angle).

\section{Discussions}

ng (1998) further demonstrated that the phase shift
between the potential and density spirals is intimately related to
the {\em gradient} of the so-called
{\em torque coupling integrals} (which is the same thing as
an angular momentum flux in the radial direction) defined in LBK,
such that during the linear modal growth process, $T(r)=-dC_g/dr$
where $C_g$ is the gravitational torque coupling integral;
and that at the quasi steady state of the wave mode, $T(r)=-(dC_a+dC_g)/dr
\equiv -(dC)/dr$
where $C_a$ is the advective torque coupling integral.
Since $T(r) <0$ inside corotation and $T(r) > 0$ outside corotation
due to the sign of the phase shift, it follows that $dC/dr>0$
inside corotation and $dC/dr<0$ outside corotation, which means
that the C(r) function is of a characteristic bell shape with
the peak of the bell at the corotation radius.  This bell shaped
angular momentum flux says that a spiral mode not only transports
angular momentum outward from the innermost region to the outer
part of the disk, as originally stated in LBK, it also picks up
angular momentum from all galactic radii inside corotation, and dumps
angular momentum unto all radii outside corotation {\em en route}
of the outward angular momentum transport.

This bell-shaped torque couple demonstrated in Zhang (1998)
turns out to be intimately related to the ability of the wave
to spontaneously grow in the linear regime, and for the
inevitability of the basic state evolution at the quasi-steady state
of the wave mode, as we show below.
Since $dC/dr=d(C_a+C_g)/dr=-dL/dt$
(i.e. the gradient of the radial angular momentum
flux is the rate of angular momentum change in the local annulus,
which is again a direct consequence of angular momentum conservation),
we have in general
\begin{equation}
{ {dC} \over {dr}} =
-  {{d (L_{basic~state} +L_{wave})} \over {dt}}
\end{equation}
based on the angular momentum conservation.

In the linear regime:
\begin{equation}
{{d L_{basic~state}} \over {dt}} = 0,
\end{equation}
therefore
\begin{equation}
-  {{dC} \over {dr} } = {{dL_{wave}} \over {dt}} =
2 \gamma_{g} L_{wave}
,
\end{equation}
where $\gamma_g$ is the amplitude growth rate of the wave mode.

At the quasi-steady state:
\begin{equation}
{{d L_{wave}} \over {dt}} = 0,
\end{equation}
therefore
\begin{equation}
-  {{dC} \over {dr} } = {{dL_{basic~state}} \over {dt}}
.
\end{equation}

We see from the above expressions that instead of the
outward angular momentum flux $C$
(as originally thought by LBK) it is rather the {\em gradient}
of this transport $dC/dr$ (which is itself proportional
to the phase shift-induced torque $T(r)$) that is responsible for
the {\em homogeneous} modal growth across the entire disk surface
in the linear regime, and for the {\em homogeneous}
evolution of the basic state at the quasi-steady state of the wave mode.

In the past discussion of secular evolution mechanisms
emphasis has been placed on the accretion of gas
under the influence of the central bar.  However, the microscopic
viscosity in the gas component is known to be inadequate to support
a reasonable accretion rate even for proto-stellar accretion
disks.  Furthermore, since gravity cannot really distinguish
whether the underlying matter is made of stars or gas,
the two component are expected to play essentially {\em identical} 
roles in the spiral-or-bar-induced viscous accretion processes.
Indeed, the two-component N-body simulations involving
both the stars and gas have invariably found that the
phase shift between the stellar and gaseous densities are
very small (Carlberg \& Freedman 1985; Zhang 1998).
It is their common phase shift with respect to the spiral
potential that caused the stars and gas to both drift towards
the center(Zhang 1998).

\section{Conclusions}

We have shown that a globally self-consistent solution for a
spontaneous spiral mode in the disk geometry can be obtained
as a dynamical equilibrium state, with the growth tendency
of the spiral mode balanced by the local dissipation in the
basic state.  The resulting secular energy and angular momentum exchange
between the wave mode and the basic state is mediated by a
temporary local gravitational instability at the spiral arms.

The closed form equations
for the rate of energy and angular momentum exchange between
the basic state and the wave mode have been quantitatively
confirmed in N-body simulations, and these expressions can be
used to compute the evolution rate of physical galaxies.
The gravitational torque mechanism is expected to be
operating in other types of astrophysical disks as well.


\begin{references}
\reference Butcher, H., \& Oemler, A. Jr. 1978, \apj, 219, 18; \apj, 226, 559

\reference Carlberg, R. G., \& Freedman, W. L. 1985, \apj, 298, 486

\reference Lynden-Bell, D., \& Kalnajs, A. J. 1972, \mnras, 157, 1

\reference Zhang, X. 1996, \apj, 457, 125;
1998, \apj, 499, 93; 1999, \apj, 518, 613
\end{references}
\end{document}